\documentclass[a4paper,twoside]{article}

\usepackage{subfigure}
\usepackage{calc}
\usepackage{amssymb}
\usepackage{amstext}
\usepackage{amsmath}
\usepackage{amsthm}
\usepackage{multicol}
\usepackage{pslatex}

\usepackage{graphicx}
\usepackage{url}
\usepackage[utf8]{inputenc}
\usepackage{listings}
\usepackage[T1]{fontenc} 
\usepackage{microtype} 
\usepackage{type1cm} 
\usepackage{siunitx}
\usepackage[english]{babel}

\usepackage{etoolbox}

\usepackage{url}
\usepackage[table,dvipsnames]{xcolor}
\usepackage{tabularx}
\usepackage{framed}
\usepackage{booktabs}
\usepackage[bookmarks=false,hidelinks]{hyperref}

\usepackage{apalike}
\usepackage{SCITEPRESS}     

\lstdefinelanguage{SpecElektra}{
        basicstyle=\ttfamily,
        comment=[l]{;},
        commentstyle=\color{purple}\ttfamily,
        breaklines=false,
        morestring=[b]',
        morestring=[b]`,
        morestring=[b]",
        stringstyle=\color{red}\ttfamily,
        sensitive=f,
        keywordstyle=\color{BlueViolet}\bfseries,
        keywordstyle=[2]\color{Melon}\bfseries,
        keywordstyle=[3]\color{Aquamarine}\bfseries\textit,
        keywordstyle=[4]\color{NavyBlue}\bfseries,
        keywordstyle=[5]\color{Mahogany},
        keywords={context, property1, property2, file, content},
        keywords=[2]{layer},
        keywords=[3]{name, interface, network, emphasized},
        keywords=[4]{[, ]},
        keywords=[5]{getenv, open, setenv},
        literate={=}{{{\color{red}\textbf=}}}1
                 {<-}{{{\color{red}\textbf<-}}}2
                 {:=}{{{\color{red}\textbf:=}}}2
                 {\%}{{{\color{NavyBlue}\textbf\%}}}1
                 {*}{{{\color{green}\textbf*}}}1
                 {[}{{{\color{Sepia}\textbf[}}}1
                 {]}{{{\color{Sepia}\textbf]}}}1,
}

\makeatletter
\def\addToLiterate#1{\edef\lst@literate{\unexpanded\expandafter{\lst@literate}\unexpanded{#1}}}
\lst@Key{add to literate}{}{\addToLiterate{#1}}
\makeatother

\lstMakeShortInline|

\rowcolors{1}{white}{lightgray}
\newcolumntype{R}{>{\raggedleft\arraybackslash}X}%

\newcommand{\blap}[1]{{\begin{tabular}[t]{@{}c@{}}#1\end{tabular}}}

\makeatletter
\newcommand{\p}[1]{\SI{#1}{\percent}}
\let\@p\@@p
\makeatother

\makeatletter
\newcommand{\property}[1]{\@property{#1}}
\newcommand{\@@property}[1]{\textsc{\let\@property\@@@property#1}}
\newcommand{\@@@property}[1]{\textnormal{\let\@property\@@property#1}}
\let\@property\@@property
\makeatother

\makeatletter
\newcommand{\keyname}[1]{\@keyname{#1}}
\newcommand{\@@keyname}[1]{\texttt{\let\@keyname\@@@keyname#1}}
\newcommand{\@@@keyname}[1]{\textnormal{\let\@keyname\@@keyname#1}}
\let\@keyname\@@keyname
\makeatother

\makeatletter
\newcommand{\strong}[1]{\@strong{#1}}
\newcommand{\@@strong}[1]{\noindent \textbf{\let\@strong\@@@strong#1}}
\newcommand{\@@@strong}[1]{\textnormal{\let\@strong\@@strong#1}}
\let\@strong\@@strong
\makeatother

\makeatletter
\newcommand{\RQ}[1]{\@RQ{RQ~#1}}
\newcommand{\@@RQ}[1]{\noindent \textbf{\let\@RQ\@@@RQ#1}}
\newcommand{\@@@RQ}[1]{\textnormal{\let\@RQ\@@RQ#1}}
\let\@RQ\@@RQ
\makeatother

\makeatletter
\newcommand{\contribution}[1]{\@contribution{\textbf{Contribution~#1:}}}
\newcommand{\@@contribution}[1]{\noindent {\let\@contribution\@@@contribution#1}}
\newcommand{\@@@contribution}[1]{\textnormal{\let\@contribution\@@contribution#1}}
\let\@contribution\@@contribution
\makeatother

\makeatletter
\newcommand{\finding}[1]{\@finding{Finding~#1}}
\newcommand{\@@finding}[1]{\noindent \textbf{\let\@finding\@@@finding#1}}
\newcommand{\@@@finding}[1]{\textnormal{\let\@finding\@@finding#1}}
\let\@finding\@@finding
\makeatother

\begin{document}

\title{Introducing Context Awareness \\ in Unmodified, Context-unaware Software}

\author{\authorname{Markus Raab\sup{1} and Gergö Barany\sup{2}\thanks{This
work was performed while the author was at CEA LIST Software Reliability
Laboratory, France, and supported by the French National Research Agency
(ANR), project AnaStaSec, ANR-14-CE28-0014.}}
\affiliation{\sup{1}Institute of Computer Languages, Vienna University of Technology, Vienna, Austria}
\affiliation{\sup{2}Inria Paris, France}
\email{markus.raab@complang.tuwien.ac.at, gergo.barany@inria.fr}
}

\keywords{Context-aware Software Engineering, Configuration Specification}

\abstract{
Software tends to be highly configurable, but most applications are hardly context aware.
For example, a web browser provides many settings to configure printers and proxies, but nevertheless it is unable to dynamically adapt to a new workplace.
In this paper we aim to empirically demonstrate that by dynamic and automatic reconfiguration of unmodified software we can systematically introduce context awareness.
In 16 real-world applications comprising 50 million lines of code we empirically investigate which of the 2,683 run-time configuration accesses
(1) already take context into account, or
(2) can be manipulated at run-time to do so.
The results show that context awareness can be exploited far beyond the developers' initial intentions.
Our tool Elektra dynamically intercepts the run-time configuration accesses and replaces them with a context aware implementation.
Users only need to specify contexts and add context sensors to make use of this potential. 
}

\onecolumn \maketitle \normalsize \vfill

\section{Introduction}
\label{introduction}

\emph{Context}---information about the environment in which software executes---strongly influences the behavior we expect from software, and most software is subject to context.
As our running example, we describe a web browser with its local network settings as context:
In different networks, web browsers may require different proxy settings for Internet access.
The default printer might also have to be changed to a physically co-located one.

If software (more) readily adapts its behavior automatically to its current context, we call it (more) \emph{context aware}~\cite{alegre2016engineering}.
Context awareness fundamentally increases user experience~\cite{dey2000towards}.
For example, if a web browser considers its network context, users will be able to display a web page regardless of which proxy is required by the network.

\emph{Context-oriented software engineering} (COSE) puts context awareness in its focus.
Previous COSE approaches required developers to consider every context already at design time~\cite{kamina2014context}.
Thus they were not applicable for already existing large software projects and in particular for legacy software.

To improve on these issues, we propose to move COSE to deployment time.
This way we delay decisions about supported contexts. 
Three classes of stakeholders participate:
(1) the developers, who still focus on configurability without the need to explicitly implement context awareness; 
(2) the administrators, who enable context-awareness in applications with our novel COSE process during deployment; and
(3) the end users, who enjoy more context-aware applications.

\subsection{Research Questions}

We claim that it is possible and practical to take unmodified software, and by run-time reconfiguration, improve their context awareness.
Our goal is to exploit already existing \emph{run-time configuration accesses} (RCAs) in free and open source software (FLOSS).

\vspace{1em}
\noindent
To validate our claim, we answer 3 research questions:

First we need to show that enough RCAs are present in FLOSS to support run-time reconfiguration.
To confirm that, in Section~\ref{source} we analyze the source code in a sample of 16 popular and large-scale FLOSS applications and count RCAs to answer
\RQ{1:} How often are RCAs used in FLOSS?

To find out whether RCAs occur sufficiently frequently during run time, a dynamic analysis is needed.
In Section~\ref{runtime} we evaluate case studies with the same 16 applications and answer
\RQ{2:} How many RCAs can be made context aware?

In Section \ref{evaluation} we investigate if our proposed solution is efficient enough by answering
\RQ{3:} What is the overhead of context-aware RCAs?

\subsection{Contributions}

This paper is (to the best of our knowledge) the first endeavor to empirically investigate context awareness in large-scale FLOSS applications: 
\begin{itemize}
\item We collected profound evidence that RCAs in FLOSS applications can be used to improve context awareness.
\item In case studies with 16 real-world applications we found out that COSE improves unmodified FLOSS applications.
\item No previous evaluation of context-aware applications was conducted using such large, complex, and popular applications.
\end{itemize}

\contribution{1}
In a source-code analysis of 16 FLOSS applications we observe that a particular kind of RCAs, namely invocation of the |getenv| function, is used pervasively (2,683 call sites).
(Section \ref{source})

\contribution{2}
We confirm that RCAs are used ubiquitously also at run time.
We systematically investigated which RCAs can be used to improve context awareness in all 16 applications.
We improved context awareness in nearly every studied application and found promising candidates in the others.
For example, from 316 candidates in browsers, in total we found 40 RCAs that certainly enable context awareness.
In some cases applications were made completely aware of individual contexts.
Furthermore, we could integrate all 1,957 configurations settings of Firefox, which provided seamless adaption to workplaces.
We never needed to modify the source code.
(Section \ref{runtime})

\contribution{3}
By evaluating performance characteristics of browsers in realistic proxy transitions, we found that in minimalist applications there is significant overhead.
For feature-rich applications such as Firefox, however, the overhead of our tool is below \p{1}.
(Section \ref{evaluation})

\section{Preliminaries \& Motivation}
\label{background}

An important aspect of software configuration is to specify \emph{run-time configuration accesses} (RCAs).
RCAs are the places within code that define different behavior based on configuration.
Usually RCAs are calls to configuration \emph{application programming interfaces} (APIs) such as |getenv| but can also be direct accesses to data structures.

The |getenv| function is a low-level configuration RCA.
It accesses the \emph{environment variables} of the current process, which is set by the caller (e.\,g., the user's shell) when a program is started.
After the start of the process, its environment variables can no longer be changed externally, only from within the process using the |setenv| function.
We chose it for most of our investigations because it is widely standardized and available in many programming languages.

\noindent
For example, in web browsers we find code such as:

\begin{lstlisting}
getenv ("http_proxy");
\end{lstlisting}

In this example, the return value of |getenv| can contain an outdated proxy after network changes because environment variables are not updated for processes.

\emph{Context-oriented programming} (COP) allows developers to naturally separate multi-dimen\-sional concerns~\cite{dey2000towards,salvaneschi2012context,schippers2010contextoperational}.
COP is one way to specify programs that adapt their behavior to the context.

\emph{Layers} are the foundation of COP~\cite{appeltauer2009contextcomparision,costanza2006efficientlayeractivation,lowis2007contextbeyond,wasty2010contextlua}.
Every layer constitutes one dimension of context that cuts across the software system.
The (de)activation of layers occurs during program execution.
All active layers together define the current context of the program.

\emph{Contextual values} originate from Lisp systems~\cite{asirelli1979flexible}.
Tanter revived contextual values as a lightweight subset of COP.
They ``boil down to a trivial generalization of the idea of thread-local values''~\cite{tanter2008contextvalues}
and can be described as variables whose values depend on the current context.
Contextual values naturally work along with the concept of layers.

In the present paper we will interpret access to configuration settings as contextual values.
Every RCA, such as a |getenv| invocation, will be considered as reading a contextual value.

\emph{Context awareness} is a property of a program and defines the degree of context taken into account.
For every possible context a combination of layers considers necessary adaptations.
Organizing such dynamic behavioral changes requires careful engineering~\cite{salvaneschi2012context}.
Alternatively, contextual values are by design always context aware~\cite{raab2016persistent}.

\emph{Context-oriented software engineering} (COSE) provides a
``methodology that guides us to a specification of context-dependent requirements''
and a systematic mapping from context-dependent use cases to layers~\cite{kamina2014context}.

\emph{Context sensors}~\cite{dey2000towards,baldauf2007survey} are hardware and software with the main purpose of activating layers according to context changes.
We will use them as separate processes that wait for context change events~\cite{raab2016persistent}.

Yin et al.~\cite{yin2011empirical} found out that in ``a large portion (\p{46.3} to \p{61.9}) of the parameter misconfigurations'' the context was not considered.
Our goal is to explicitly specify the contextual values that make the execution of the applications more context-aware.
This specification leads to a better understanding of the context in teams maintaining the software.  
We are positive that the specification helps reduce external misconfiguration errors as a side effect.

For newly written context-aware software, current COP and COSE approaches would be a viable choice~\cite{hong2009context}.
For large FLOSS projects, however, rewriting the whole source code is not feasible.

\section{Elektra}
\label{elektra}

Elektra is a library developed by one of the authors, which implements uniform, consistent and context-aware configuration access.
In the present paper we describe an approach to use Elektra as a tool to integrate unmodified applications.
The approach is to apply Elektra in COSE processes at deployment.

Elektra~\cite{raab2016elektra} works as follows:
At application start, Elektra initializes itself by parsing configuration files.
The configuration files contain both the specifications and configurations for contextual values.
Elektra supports over 190 configuration file formats including the widely-used INI, XML and JSON formats.
The support for these formats enables Elektra to directly manipulate configuration of applications.

In its essence, Elektra provides a key-value database with unique keys and a specification for every configuration setting.

\subsection{Interception}

To work with unmodified applications, Elektra intercepts important library calls, including the following:
\begin{itemize}
\item Each |getenv| invocation to provide context-aware access for environment variables.
\item Each |open| invocation to return configuration files with configuration settings respecting the current context.
\end{itemize}

Interceptions of library calls are platform-dependent but are available for every major OS, e.\,g., |LD_PRELOAD| and |/etc/ld.so.preload| for Linux.
Instead of requiring developers to implement new behavior for context adoption, we rely on already existing behavioral adoptions that are guarded by RCAs.

\subsection{Context Specification}

Elektra itself is configured via a configuration specification language.
In this paper we will use a simple key-value syntax to illustrate the specification of the key-value database and its contents.
For example, let us specify the contextual value |getenv/http_proxy|:

\begin{lstlisting}
[getenv/http_proxy]
context=http_proxy/%interface%/%network%
\end{lstlisting}

The key within |[]| represents a unique identifier to a configuration setting.
Entries in the database are organized hierarchically with |/| as the level separator.
The |getenv|-interceptor reads its configuration from keys starting with |getenv/|.
We configure it to handle |getenv| invocations with the parameter |"http_proxy"|.

In the example above, the only property for this key, i.\,e.,~|context|,  specifies that the value to be returned from such invocations should be context aware.
Using the |
The |getenv| invocation returns the proxy configured for the currently active |interface| and |network| layers.
This functionality allows us to modify configuration settings passed to applications:
An application that requests a configuration setting from the environment transparently receives a setting from our key-value database instead.
By honoring context in the lookup we introduce context awareness in the client software.

The context-aware lookup makes sure that the returned value recursively respects context specifications.
With changing context, i.\,e., different values in layers, the same requested key has different values.
We express the possible values using straight-forward pattern matching.
E.\,g., the placeholder |*| will match any layer that was not matched specifically by name:

\begin{lstlisting}
http_proxy/wlan/home= proxy.example.org
http_proxy/eth/work = proxy.example.com
http_proxy/*/* = default.example.com
\end{lstlisting}

Personalization is an important aspect of context-aware systems~\cite{alegre2016engineering}.
In Elektra we personalize applications by changing such configuration values for every individual context.

\subsection{Context Changes and Sensors}

When the context changes, this information must be communicated to applications.
For the present work, we use external \emph{context sensors}: small programs running in separate processes that monitor the context of interest.
When the sensor detects a change, it updates the corresponding layer's value in the key-value database.
Future requests for contextual values via Elektra (through intercepted |getenv| or |open| invocations) will use these updated layer values in their lookups.
Having context sensors running in their own process separates concerns between the application and the code detecting context changes.

In the running example, users switch networks by changing their location or by connecting a network cable.
A sensor detects this change and updates Elektra's database accordingly.
We used hooks in network interfaces to implement this use case.
Assume that the |interface| changes to |eth| and the |network| to |work|.
Then the next |getenv("http_proxy")| invocation will return |proxy.example.com|.
This is an increase in context awareness:
Normally |getenv| is \emph{not} context-aware because the program's environment is initialized at startup and cannot be modified externally, only by the program itself using |setenv|.
Thus the standard |getenv| function always returns the same value for a given argument.
In contrast, Elektra's modified |getenv| returns different values if the underlying context changes.

Elektra is not limited to pulling configuration settings while RCAs are executed.
Instead Elektra can push information to applications by notifying them to reload their configuration, e.\,g., via signals for daemons or socket communication for Firefox.

\section{RQ1: Use of \texttt{getenv}}
\label{source}

In this section we collect empirical evidence of |getenv| invocations in the source code of applications.

\subsection{Methodology}

We count the total lines of code and occurrences of |getenv| in selected applications.
Obvious wrapper functions (e.\,g., |LYGetEnv| in Lynx) are treated identically to |getenv| itself.

To improve external validity we carefully sampled 16 applications.
We started by including large applications that have a thriving community.
In addition we took care to have a broad range of diverse applications.
We searched for further popular applications to reduce the familiarity heuristic.
If Internet pages repeatedly mentioned some applications, we considered them for inclusion.
Finally, we set a focus on browsers to have a better picture for a specific domain.

The evaluation of the paper consistently uses the same applications with the same versions.
We analyzed applications in the version as included in Debian Jessie 8 amd64 available at \url{snapshot.debian.net}.
Considering these factors we compiled the following list of 16 applications and versions:

\noindent
{\small
\begin{tabularx}{\columnwidth}{l  R |  l R}
\toprule
{\bfseries application} &
{version} &
{\bfseries application} &
{version}
\\
\hline

0ad & 0.0.17                           &          Gimp & 2.8.14\\
Akonadi & 1.13.0                       &          Inkscape & 0.48.5\\
Chromium & 45.0.2454                   &          Ipe & 7.1.4\\
Curl & 7.38.0                          &          Libreoffice & 4.3.3\\
Eclipse & 3.8.1                        &          Lynx & 2.8.9dev1\\
Evolution & 3.12.9                     &          Man & 2.7.0.2\\
Firefox & 38.3.0esr                    &          Smplayer & 14.9.0~ds0\\
Gcc & 4.9.2                            &          Wget & 1.16\\

\hline

\bottomrule

\end{tabularx}
}

We used Cloc 1.60 to determine the code size of the applications in the versions as listed above.
We used  |grep -rio| to find all textual |getenv| occurrences.
Finally we manually looked at every |getenv| occurrence to check whether it is an invocation or something else like text in a comment.

\subsection{Results}

In the following table below the column \emph{1k lines of code} shows the code size of the applications, expressed as multiples of 1,000 lines of code.
For the column \emph{counted getenv} we manually counted |getenv| invocations.

{\small
\noindent
\begin{tabularx}{\columnwidth}{l  R  R  R}
\toprule
\multicolumn{1}{l}{\bfseries application} &
\multicolumn{1}{R}{\bfseries \blap{1k lines\\of code}} &
\multicolumn{1}{R}{\bfseries \blap{counted \\ getenv}} &
\multicolumn{1}{R}{\bfseries \blap{lines per\\ getenv}}
\\
\hline

0ad		&	474   	&	55 	&	8,617  \\
Akonadi 	&	37    	&	13 	&	2,863  \\
Chromium 	&	18,032	&	770	&	23,418 \\
Curl 		&	249    	&	53 	&	4,705  \\
Eclipse 	&	3,312 	&	40 	&	82,793 \\
Evolution 	&	673   	&	23 	&	29,252 \\
Gcc 		&	6,851 	&	377	&	18,172 \\
Firefox 	&	12,395	&	788	&	15,730 \\
Gimp 		&	902   	&	56 	&	16,102 \\
Inkscape 	&	480   	&	19 	&	25,255 \\
Ipe 		&	116   	&	21 	&	5,529  \\
Libreoffice 	&	5,482 	&	284	&	19,304 \\
Lynx 		&	192   	&	89 	&	2,157  \\
Man 		&	142   	&	62 	&	2,293  \\
Smplayer 	&	76   	&	1  	&	76,170 \\
Wget 		&	143   	&	32 	&	4,456  \\
\hline
Total	&	49,556	&	2,683	&	18,470 \\
Median	&	477	&	54	&	       \\

\bottomrule

\end{tabularx}
}

The applications we analyzed have 2,683 |getenv| invocations in 50 million lines of codes.
We excluded textual occurrences in wrappers, comments, ChangeLogs or similar.

\begin{framed}
\noindent

\finding{1:}
We demonstrate that |getenv| is used pervasively by finding 2,683 invocations.
This is one |getenv| occurrence in 18,470 lines of code.
\end{framed}

\section{RQ2: Run-Time Behavior}
\label{runtime}

In this section we validate the applicability of our approach at run-time.
Run-time analysis considers |getenv| invocations by all participating libraries, complementing our source-code analysis.
We will investigate how often changed return values of |getenv| invocations actually modify the application's behavior to improve context awareness.

This study is a partial replication of our previous study~\cite{raab2016unanticipated}.
Unlike the previous study, we compare the context awareness of every single parameter.
Furthermore, we added more applications and introduce |open| interception. 
For brevity, however, we only report about selected and representative cases.

\subsection{Methodology}

We applied our approach for all 16 applications.
As described in Section~\ref{elektra}, we use the library preload mechanism to use Elektra's implementations of |getenv| and |open| instead of the ones in the standard library.

First we started the 16 applications and clicked through the user interface.
While doing so, we logged every |getenv| invocation and its parameters.
To check if the |getenv| invocation is used as if standard |getenv| were context aware, we modified the return values of |getenv| while the application was running.
Then we repeated the user-interaction to see if the |getenv| invocation influences the behavior.

\subsection{Results}

\makeatletter
\preto\tabular{\global\rownum=\z@}
\makeatother

\noindent
{\small
\begin{tabularx}{\columnwidth}{ l  R  R  R  R  R }
\toprule
\multicolumn{1}{l}{\bfseries application} &
\multicolumn{1}{R}{\bfseries \blap{getenv \\ all}} &
\multicolumn{1}{R}{\bfseries \blap{all \\ uniq}} &
\multicolumn{1}{R}{\bfseries \blap{later \\ uniq}} &
\multicolumn{1}{R}{\bfseries \blap{later \\ config}} &
\multicolumn{1}{R}{\bfseries \blap{context \\  aware}}
\\
\hline
Chromium   & 2,723  & 1,056    & 73 &$\geq24$& $\geq1$  \\
Curl       & 87     & 14       & 9  &       6&       6  \\
Firefox    & 8,185  & 273      & 210&     118&$\geq15$  \\
Lynx       & 1,428  & 45       & 23 &      19&      16  \\
Wget       & 13     & 7        & 1  &       1&       1  \\
\bottomrule
\end{tabularx}
}

The table above shows the number of |getenv| invocations on a freshly installed Debian system.
The number of invocations varies widely from system to system depending on configuration and installed software.
E.\,g., on other machines, we observed up to four times more unique |getenv| invocations during run-time for Firefox.
Sometimes we found settings that are likely to be context aware (indicated by $\geq$) but lacked the resources to investigate them in detail.

In the column \emph{getenv all} we see how many times the browsers called |getenv| in total.
The next column shows the number of |getenv| invocations with unique parameters.
The column \emph{later uniq} only considers |getenv| invocations with unique parameters and only after startup.
The next column are candidates for context awareness: they are additionally related to configuration.
The last column shows which of the candidates actually successfully influenced behavior at run time without reloading.

In the analysis we found many |getenv| invocations after startup.
Loops implementing user interactions often repeatedly call functions that redo the same |getenv| invocations.

\subsection{Case Study: Firefox}

In a case study we conducted the complete COSE process.
We selected Firefox and specified |http_proxy| and |PRINTER_LIST| as configuration options of interest as shown in Section~\ref{elektra}.
We implemented the layer-changes with one-line hooks in the |/etc/NetworkManager| scripts.

Then we needed to specify printers/proxy for every context line-by-line.
Within a day, Firefox fully-automatically selected nearby printers and proxies immediately on network changes (available printers are even modified while the printer dialog is open).

Elektra also allows us to modify options in configuration files.
We needed 9 hours to configure Firefox to enable rereading its configuration files.
In 2 more hours we implemented an Elektra plugin for Firefox's configuration files.
With |open| interception we have a context-aware mechanism for all of 1,957  configuration options available in Firefox's configuration files~\cite{jin2014configurations}. 

\begin{framed}
\noindent

\finding{2:}
In each of the 16 application user interactions caused |getenv| invocations, often useful to make features flawlessly context aware.

We successfully used Elektra in a real-world case study with Firefox.
To enable our implementation of retrofitting context-awareness for flexible workplaces, only three actions were required:
(1) specify contextual values,
(2) create context mapping for every workplace, and
(3) add context sensors to switch layers.
\end{framed}

\strong{Implication:} Our tool can be practically applied in real-world case studies with small effort.

\section{RQ3: Performance Evaluation}
\label{evaluation}

To evaluate the performance we profiled different browsers during a proxy transition on a hp\textsuperscript{\textregistered} EliteBook 8570w.
In the experiment we opened a web page, then changed the context, and finally opened a different web page.
For the proxy transition Elektra performs the following steps for every process:
First the process needs to parse the context specification.
Then the actual |getenv| is replaced with our context-aware implementation.
On layer transitions, the configuration file needs to be reread.

We measured the number of executed CPU instructions with Valgrind's tool Callgrind. 
We report inclusive costs, i.\,e., the cost of the |getenv| invocation including every callee.
Thus Valgrind simulates a CPU, the results are deterministic.

\strong{Results:}
In the first benchmark we will use the Lynx browser.
It is written in a lean way and has negligible startup times.
Such an efficient implementation allows more precise exploration of the impact the context switches have.

First we started Lynx without Elektra and visited two links.
Valgrind counted 92,888,073 instructions (median of three invocations).
Then when we activated Elektra and changed the proxy before visiting the second link, we counted 114,049,336 instructions, which are about \p{18.5} more instructions.
We used the context specification with the two layers |network| and |interface|.

Without context-aware |getenv|, the |getenv| invocations needed \p{0.33} of all instructions.
If we modify |getenv/http_proxy| directly (without using layers) |getenv| needs \p{24.51}.
If using the setup with the two layers, |getenv| invocations needed \p{25.27}. 

With Firefox the comparison was more difficult because it consumes resources even without user interaction.
The startup times with Valgrind are nearly two minutes.
We estimated the overhead by looking at the profile data, similar to the |getenv| overhead in Lynx.
Overall 20,362,848,539 instructions were needed to display two web pages.
Internal Elektra overhead, by summing up all costs from the Elektra library, are 68,750,481, i.\,e., \p{0.39}. 
The |g_getenv| function (a wrapper for |getenv| used within Firefox) needs 16,614,089 instructions (i.\,e., \p{0.08}) instead of 22,703 instructions (i.\,e., \p{0.00}) without Elektra.

\begin{framed}
\noindent

\finding{3:}
In minimalist applications such as Lynx our approach can cause some overhead.
The number of participating layers caused only minimal differences.
For feature-rich applications the overhead is below \p{1}.
\end{framed}

\section{Related Work}
\label{related}

Xu et al. investigated which configuration settings are actually used in practice~\cite{xu2015hey}.
They argue that users get confused by too many settings.
We fully agree and think that our approach helps here by automatically deducing most settings from context.
Then developers can remove these settings from user guides.
Advanced users, however, still can override context-aware configuration settings.

Jin et al. describe different challenges in configuring real-world systems~\cite{jin2014configurations}. 
Our approach addresses them by (1) working across language barriers, (2) having a holistic integration of different RCAs, and (3) making sure that RCAs do not return outdated values.
The authors uncovered 1,957 settings for Firefox and assumed that only a ``small part'' of settings is missing.
But our study shows, that even |getenv| alone adds a large amount of otherwise unconsidered configuration settings.

Most other approaches require modifications in the source code.
Tanter et al.~\cite{tanter2006context} propose to make aspects context aware.
COP~\cite{salvaneschi2012context,appeltauer2009contextcomparision,lowis2007contextbeyond,baldauf2007survey,hong2009context,raab2015global} improves the modularity in programs.
In earlier work we investigated how contextual values are synthesized with code generation~\cite{raab2015global,raab2014program,raab2015kps}.
A survey discusses many different approaches how to implement context-aware applications~\cite{alegre2016engineering}.
Mens et al.\ created a taxonomy for context-aware variability approaches~\cite{mens2016taxonomy}.
All these approaches require at least some context-specific design upfront the implementation.
The specification language Elektra does not only improve context awareness, but fosters system integration~\cite{raab2016improving}.
Some approaches focus on deployment, but require decisions at design time~\cite{lee2014deployment}.
Alexandrov et al. facilitates intercepting of library calls to improve user experience, but with a different goal~\cite{alexandrov1998ufo}.

The survey of Xu and Zhou gives an overview of the different approaches for improving on configuration problems~\cite{xu2015systems}.
In contrast, our basic idea is to automatically derive correct configuration settings from context.

\section{Threats to Validity}

\paragraph*{Internal:}
Both the code and run-time analysis have the danger of subjective classification and oversight.
To minimize such errors we included second opinions and only report large differences.
Additionally the combination of code and run-time analyses yields a more complete picture as proposed for mixed methods~\cite{ihantola2011threats}.

\paragraph*{External:}
An important concern is whether the evaluated applications and their developers are representative.
We address it by studying a high number of diverse applications.
We included both small and large applications.
We took care that different domains, development teams and programming languages are represented.
In particular the browsers are used heavily in mobile contexts.

We have to acknowledge that most software we evaluated is written in C/C++.
Nevertheless, Java, JavaScript and Python were well represented with 4.3, 3.3, and 1.1 million lines of code, respectively.
Furthermore, we added Eclipse to also have a large project mainly implemented in Java.
Since we found no context-aware application of reasonable size with an active community,
we could not include such applications into the analysis.
Hence our claims exclude applications developed with context-awareness as goal.

An equally important concern is whether |getenv|, our main subject of study, represents every form of RCA.
Based on many previous studies~\cite{jin2014configurations,rabkin2011static,xu2013blame}, run-time RCAs are in their essence simple key-value accesses.
Higher-level RCAs, e.\,g. with type-safety, would only complicate the implementation.

Because we did an in-depth source code analysis, we could not pick closed-source applications.
A significant portion of the evaluated software, however, has at least roots as closed-source applications.
Also based on experience within companies, we are positive that our conclusions hold for closed-source applications.

It is well known that in experimental analysis high standards are required~\cite{johnson2002theoretician};
e.\,g., to mitigate measurement issues we always used two different profiling tools.

Overall we cannot draw any general conclusions applicable to every form of configuration.
In particular we focus on observations how RCAs are used in FLOSS applications.
Thus results should be interpreted and generalized with awareness of our focus.
Nevertheless our study provides profound insights about connections between configuration and context awareness.

\section{Conclusion and Future Work}
\label{conclusion}

In this paper we claimed that unmodified applications can become more context aware.
We demonstrated that such an approach exists and is practical.
We evaluated the approach on 16 large, real-world FLOSS applications.
By configuring a simple tool, context awareness was improved in case studies, often even flawlessly.
We applied a straightforward context-oriented software engineering process which enables systematic applicability during deployment.

Our work shows that it is realistic to deduce configuration settings from context.
We are positive that doing so contributes to reduce one of the major source of configuration errors, i.\,e., forgetting about context in configuration.

We propose that environment variables should be specified and documented like other configuration settings.
Our approach shows that it is not necessary that developers foresee every possible context.
Instead layers and configuration settings per context are introduced during deployment.
Our approach is modular because context sensors are implemented separately from applications.

\noindent
Elektra is available as free software from
\par \hspace{2em} \url{http://www.libelektra.org} \par \noindent
and is more general than described in this paper:
For example, it can be used for newly developed context-aware software by generating contextual values.
This paper focuses on its use for unmodified software.

The source code analysis suggest that dependency injection (`hijacking' existing |getenv|/|open| invocations or other APIs) makes it easy to introduce context awareness.
Elektra is not limited to intercepting |getenv| and |open|.
For example, we implemented the |gsettings| API which has the potential to make GNOME settings context-aware.
As future work Elektra can be extended to make even more forms of configuration context-aware (configuration for modules, plugins, dependency injections, rich mobile APIs etc.).

Although we did not find a single occurrence where existing context awareness conflicted with our approach, combining Elektra with already context-aware software is future work.
Another research direction is to investigate how many applications allow reloading of configuration settings.
Enabling Elektra to push configuration settings to more applications improves the user experience again.

\bibliographystyle{apalike}

\end{document}